\documentclass[%
 aip,
 amsmath,amssymb,
 reprint,%
]{revtex4-1}

\usepackage{graphicx}% Include figure files
\usepackage{dcolumn}% Align table columns on decimal point
\usepackage{bm}% bold math
\usepackage[utf8]{inputenc}
\usepackage[T1]{fontenc}
\usepackage{mathptmx}
\usepackage{etoolbox}
\usepackage{bbm}
\usepackage{xcolor}
\usepackage{hyperref}
\makeatletter
\def\@email#1#2{%
 \endgroup
 \patchcmd{\titleblock@produce}
  {\frontmatter@RRAPformat}
  {\frontmatter@RRAPformat{\produce@RRAP{*#1\href{mailto:#2}{#2}}}\frontmatter@RRAPformat}
  {}{}
}%
\makeatother
\begin{document}

\preprint{AIP/123-QED}

\title[Active-Absorbing Phase Transitions in the Parallel Minority Game]{Active-Absorbing Phase Transitions in the Parallel Minority Game}
% Force line breaks with \\
\author{Aryan Tyagi}
\affiliation{
School of Computational \& Integrative Sciences,
Jawaharlal Nehru University, New Delhi-110067, India
}%

\author{Soumyajyoti Biswas}%

\affiliation{ 
Department of Physics, SRM University - AP, Andhra Pradesh 522240, India
}%

\author{Anirban Chakraborti}
  \email{anirban@jnu.ac.in}
  %\homepage{https://www.jnu.ac.in/Faculty/anirban/index.html.}
\affiliation{
School of Computational \& Integrative Sciences,
Jawaharlal Nehru University, New Delhi-110067, India
}%

\date{\today}% It is always \today, today,

\begin{abstract}
The Parallel Minority Game (PMG) is a synchronous adaptive multi-agent model that
exhibits active-absorbing transitions characteristic of non-equilibrium statistical
systems. We perform a comprehensive numerical study of the PMG
under two families of microscopic decision rules: (i) agents update their choices based on instantaneous population in their alternative choices,
and (ii) threshold-based activation that activates agents movement only
after overcrowding density crossing a threshold. We measure time-dependent and steady state limits of activity \(A(t)\), overcrowding fraction \(F(t)\) as functions of the control parameter
\(g=N/D\), where \(N\) is the number of agents and \(D\) is the total number of sites. Instantaneous rules display mean-field directed-percolation (MF-DP)
scaling with \(\beta\approx1.00\), \(\delta\approx0.5\), and \(\nu_{\parallel}\approx2.0\). Threshold
rules, however, produce a distinct non-mean-field universality class with
\(\beta\approx0.75\) and a systematic failure of dynamical scaling. We
show that thresholding acts as a relevant perturbation to the critical behavior of the model. The results highlight how minimal cognitive features at the agent level
fundamentally alter large-scale critical behavior in socio-economic and active
systems. 
\end{abstract}

\maketitle

\begin{quotation}
Collective behavior in adaptive multi-agent systems often displays emergent macroscopic
phenomena that are qualitatively distinct from the microscopic decision rules of
individual agents, a theme central to both statistical physics and socio-economic
modeling\cite{chakrabarti2006econophysics,chakrabarti2023data,sociophysics2014}. A paradigmatic example is the Minority Game (MG), introduced as a minimal model
of competitive decision-making under bounded rationality
\cite{arthur1994,challet1997,challet1998,challet1999}, and subsequently developed into a
general framework for studying coordination, resource allocation, and market-like
dynamics \cite{challet2005,chakraborti2015,chakraborti2011econophysicsII}. Extensions of the MG, including spatial and
local variants, have revealed rich collective behavior arising from interaction topology
and information structure \cite{challet2001local,moelbert2002local,huang2012}. Minority-game-inspired models have been applied to wealth inequality, herding control,
and social efficiency in competitive environments
\cite{ho2004,zhang2016,bottazzi2007,dhar2011,chakrabarti2009,ghosh2010,biswas2012}.  
Synchronously updated systems that admit absorbing states (dynamics ceases entirely), known as active-absorbing transitions, are a cornerstone of
non-equilibrium statistical mechanics, with directed percolation (DP) emerging as the
generic universality class under broad conditions
\cite{henkel2000,odor2004,lubeck2004,henkel2008}. Recently, the Parallel
Minority Game (PMG) has been introduced as a multi-choice, globally coupled extension of
the MG, in which agents switch synchronously between two fixed options drawn from a larger
set \cite{biswas2021,vemula2025}. The resulting non-conserved population dynamics and shared
resource constraints naturally generate absorbing states and critical behavior.
Understanding how adaptation mechanisms influences the system efficiency becomes very important\cite{sysiaho2004adaptive}. Here, the contrast between instantaneous updates and threshold-based activation not only determines the universality class of the transition but also offers a controlled framework to probe how minimal cognitive constraints function as relevant perturbations in non-equilibrium socio-economic systems.
\end{quotation}

\section{Introduction}
% =====================================================================
Competitive multi-agent systems often exhibit emergent macroscopic behavior that is
qualitatively different from individual decision rules. The Minority Game (MG) framework,
introduced by Challet and Zhang~ \cite{challet1997}, provides a minimal model of such
competition: agents make binary choices and the minority side wins. The original MG and its variants, including the local MG where agents interact with a subset of the population, have been extensively studied for their phase transitions and collective behaviors~\cite{challet2001local}. In particular, the Local Minority Game introduced by Moelbert and De Los Rios considers agents placed on a lattice interacting with their nearest neighbors, leading to spatial correlations and an annealed disorder that differs from the quenched disorder of the original MG~\cite{moelbert2002local}. MG variants have
been used to model financial markets, resource allocation, and collective dynamics in
complex adaptive systems~\cite{challet1999,chakraborti2015}. The Parallel Minority Game (PMG), as detailed in recent extensions~\cite{biswas2021,vemula2025,pmg_vac}, generalizes the MG to scenarios with \(D > 2\) sites, where each agent is restricted to switching between only two fixed choices among the \(D\) available sites.
The objective for the players remains the same viz., to be in the minority among the two available choices to them. But with $D>2$, each site can now contain agents having their alternate choice distributed among the various other sites. 
This overlap in choices creates coupled, parallel instances of the MG, introducing non-conserved population dynamics and heightened competition due to shared resources across agent groups.

The synchronous (parallel) update in PMG introduces non-equilibrium dynamics,
including absorbing states where the system freezes and activity ceases. Absorbing-state
transitions are central to non-equilibrium statistical mechanics; showing robust universality under broad conditions \cite{henkel2008,henkel2000,odor2004}.
(see Henkel et al.~\cite{henkel2008} and Hinrichsen~\cite{henkel2000}). In PMG, the frozen state, where no agents switch due to balanced populations, serves as the absorbing state, with global interactions approximating mean-field behavior.

In this study, we explore how microscopic decision rules shape the universality class of
the PMG. In particular, we compare instantaneous decision rules with
threshold-based rules. Our empirical finding is that
instantaneous rules map to the expected mean field behavior, while threshold rules shift the system to a distinct
universality class. This is consistent with earlier work on adaptive minority games, where threshold-based decision mechanisms were found to significantly alter system behavior and drive it toward different regimes~\cite{sysiaho2004adaptive}.

% =====================================================================
\section{Model, Strategies and Observables}
\label{sec:model}
% =====================================================================
We consider \(N\) agents distributed randomly and uniformly among \(D\) sites and a second alternative site is also assigned to each agent randomly. Each agent is restricted to move between exactly the two fixed choices for them (e.g., agent \(i\) toggles between sites \(x_i\) and \(y_i\)), and every site is ensured to be at least of the choices of at least one of the agents i.e., there is no site that is not part of the game at all. The control parameter is
\[
g=\frac{N}{D},
\]
representing the average population density of the choices. All simulations are performed for $D=500$, and the value of $N$ is fixed when a particular $g$ value is chosen. 
The point of interest is the behavior of the system near \(g=g_c=1\), where the average occupancy per site reaches unity, marking the onset of overcrowding and the active-absorbing transition. Below this point, on average, no agent will be in the majority location.  The random allocation of the agents' choices create overlaps in the choices (any given choice/site has agents whose alternate choice can be from any one of the rest $D-1$ choices). This situations create coupled minority sub-games. Each agent \(i\) at time \(t\) takes an action \(a_i(t)\in\{0,1\}\), representing their current site choice: $a_i(t)=0$ implying no change in choice for the next step and $a_i(t)=1$ implying a switch to the alternate choice in the next step.

This model has been studied before~\cite{biswas2021} in the context of movement of infected agents when an infectious disease is spreading in different locations, and also subsequently in the context of finding efficient strategy for the agents to be in the minority in the limit of high population density ($g$)~\cite{vemula2025}. Of course, for high $g$ values, some agents will always be in the overcrowded sites, but a memory-based efficient strategy could maximize (within the strategies explored) the fraction of agents in the minority. 

In this work, however, we look at the low density phase ($g$ close to one), where it is possible for $g\le 1$ to reach a state where no agent is in overcrowded sites. But for $g>1$, the overcrowding can be represented by an active-absorbing transition.
For characterization of such a transition, we then define the observables activity and overcrowding fraction, which quantify overcrowding and persistent competition in the model:

\begin{align}
A(t)&=\frac{1}{D}\sum_{s=1}^D \max\big(n_s(t)-1,0\big),\\
F(t)&=\frac{1}{D}\sum_{s=1}^D \mathbbm{1} \{n_s(t)>1\},
\end{align}

\noindent where \(n_s(t)\) is the occupancy at site \(s\), and \(\mathbbm{1} \{\cdot\}\) is the indicator function. %In the absorbing phase (\(g < g_c\)), both \(A(t)\) and \(F(t)\) decay to zero, signaling efficient resource allocation with at most one agent per site. In the active phase (\(g > g_c\)), sustained fluctuations persist. These observables align with standard order parameters for absorbing transitions~\cite{henkel2008}.

\subsection{Stochastic Strategy Formulation}
For comparative analysis, we consider two stochastic strategies adapted from recent work on efficient PMG dynamics~\cite{vemula2025}:

\paragraph{Strategy A: Instantaneous minority comparison}
Agents use current population information from both choices and switches with a probability given below, only if \(n_{x_i(t)} > n_{y_i(t)}\) and $n_{x_i(t)}>1$:
\[
p_{i}^{A}(t) = 
\frac{n_{x_i(t)} - n_{y_i(t)}}{2n_{x_i(t)}}
\]
where \(y_i(t)\) is the alternate location. $p_{i}^{A}(t)=0$, if $n_{x_i(t)} \le n_{y_i(t)}$ or $n_{x_i(t)}\le 1$. This strategy directly implements minority seeking with current information, similar to memory-less updates.

\paragraph{Strategy B: Threshold comparison}
Agents switch with a probability only if there is a positive deviation from the global average occupancy ($n_{x_i}(t)>g$ and $n_{x_i}(t)>1$):
\[
p_{i}^{B}(t) = \frac{ n_{x_i}(t)-g}{2n_{x_i}(t)},
\]
where \(x_i(t)\) is the current location of agent \(i\), \(n_{x_i}(t)\) is the population at that location, and \(g = N/D\) is the global average occupancy. $p_{i}^{B}(t)=0$, if $n_{x_i}(t)\le g$ or $n_{x_i}(t)\le 1$. This strategy drives population toward uniform distribution but doesn't consider minority status while making the decision on switching.

% =====================================================================
\section{Scaling Theory and Static Critical Behavior}
\label{sec:scaling}
% =====================================================================
%Directed percolation (DP) represents the paradigmatic universality class for non-equilibrium
%phase transitions into absorbing states~\cite{henkel2008,henkel2000}. The DP conjecture suggests that any system exhibiting
%(i) a unique absorbing state, (ii) short-range interactions, (iii) no additional conservation laws,
%and (iv) Markovian dynamics should belong to the DP universality class~\cite{odor2004}. For mean-field DP, the critical exponents are well-established: the order parameter exponent
%\(\beta \approx 1\), the temporal decay exponent \(\delta \approx 1\), and the temporal correlation length exponent
%\(\nu_{\parallel} \approx 1\)~\cite{henkel2000}. The Parallel Minority Game with instantaneous decision rules
%satisfies all DP conditions: the frozen state where no agents switch constitutes an absorbing
%state, interactions are global but mean-field-like (due to overlaps), no conservation laws constrain the dynamics,
%and instantaneous updates ensure Markovian evolution~\cite{biswas2021}.
The objective of the players in the Parallel Minority Game, as mentioned before, is to be in the minority between their two choices. The maximum utilization, therefore, is achieved when the fluctuation in the population among the different choices is low (same measure as the classical Minority Game problem, where there are only two choices). This assumes, of course, that all choices are nominally similar in terms of the number of player allocated, which is true since the players are assigned their choices randomly and uniformly (held fixed throughout the game). Now, when the number of players equal the number of choices (i.e., $g=1$), a situation would arise where each choice is occupied by one agent, which is an absorbing phase, meaning either strategy would not give a non-zero probability for any agent to switch. The dynamics becomes interesting, however, when $g>1$. Due to the strategies, the actual number of players who would now be in overcrowded locations (having population greater than one), can now be more than what is the minimum required ($gN-D$). We quantify the scaling behavior of this population dynamics, using the definitions of activity and overcrowding fractions defined above, in the steady state limit, to explore the critical behavior. 

 We probe the steady-state values \(A_{\infty}(g)\) and \(F_{\infty}(g)\) for \(g>g_c\)
and extract the order-parameter exponent \(\beta\) via
\[
A_{\infty} \sim (g_c - g)^\beta, \qquad F_{\infty} \sim (g_c - g)^\beta.
\]
The finite-size scaling analysis requires careful treatment of system sizes and fitting ranges
to avoid crossover effects near \(g_c\). We employ ensemble averaging over multiple realizations to ensure statistical reliability.

% Static scaling figure - 2 rows and 2 columns
\begin{figure*}[]
\centering
  \includegraphics[width=0.45\linewidth]{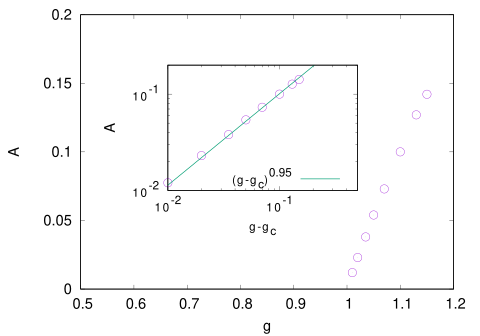}
  \includegraphics[width=0.45\linewidth]{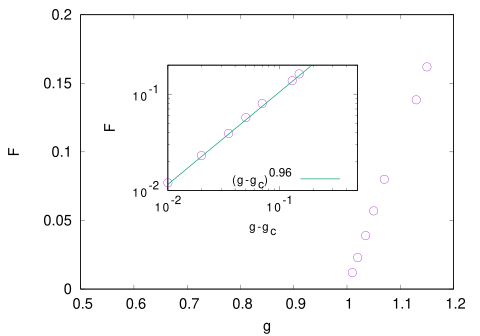}
  \includegraphics[width=0.45\linewidth]{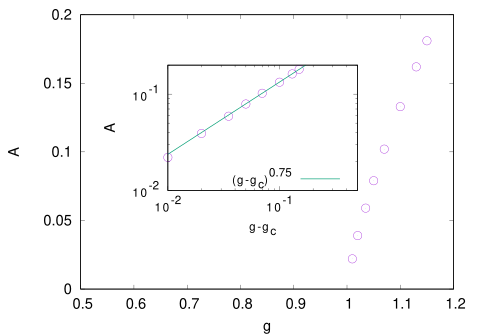}
  \includegraphics[width=0.45\linewidth]{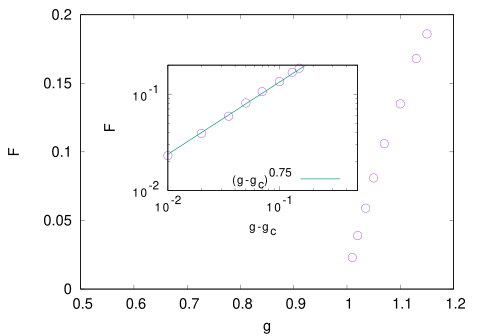}
\caption{Static scaling fits for the steady-state activity and overcrowding fraction near \(g_c=1\).  Top row: Instantaneous rules (strategy A) yield \(\beta \approx 0.95\pm0.05\), consistent with mean-field behavior . Bottom row: Threshold rules (strategy B) yield \(\beta \approx 0.75\pm0.05\), indicating a different universality class.}
\label{fig:beta_group}
\end{figure*}

The data for strategy A display near-linear behavior on a log-log plot (see Fig.~\ref{fig:beta_group}),
with \(\beta_{A}^{\rm inst}\simeq0.95\pm0.05\) and \(\beta_{F}^{\rm inst}\simeq0.96\pm0.05\),
confirming the expected mean field scaling ~\cite{henkel2008}. However, in the strategy B, the data exhibit systematically lower slopes corresponding
to \(\beta\approx0.75\) (consistently measured across both observables and different threshold
variants), indicating a clear deviation from the expected mean field universality. This aligns with earlier findings that adaptation mechanisms in minority games can drive systems toward different behavioral regimes~\cite{sysiaho2004adaptive}, where threshold-based strategies alter the fundamental dynamics.

% =====================================================================
\section{Dynamical Scaling Framework}
\label{sec:dynamics}
% =====================================================================
Dynamical scaling provides a more stringent test of universality than static exponents alone.
For systems exhibiting absorbing phase transitions, the time-dependent order parameter follows
the scaling form:
\[
A(t) = t^{-\delta}\,\mathcal{A}\!\big(t|g-g_c|^{\nu_{\parallel}}\big),
\]
where \(\delta\) is the temporal decay exponent and \(\nu_{\parallel}\) is the temporal correlation length exponent~\cite{odor2004}. Similarly, for the overcrowding fraction,
\[
F(t) = t^{-\delta}\,\mathcal{F}\!\big(t|g-g_c|^{\nu_{\parallel}}\big),
\]
with the same exponents \(\delta\) and \(\nu_{\parallel}\), and scaling function \(\mathcal{A}(x)\) sharing asymptotic behaviors with \(\mathcal{F}(x)\). For the mean field limit, the expected exponents are \(\delta_{\rm MF}=1\) and \(\nu_{\parallel,{\rm MF}}=1\).

The scaling function \(\mathcal{F}(x)\) has the asymptotic behavior: \(\mathcal{F}(x) \sim \text{constant}\)
for \(x \ll 1\) (critical quenching), and \(\mathcal{F}(x) \sim x^\beta\) for \(x \gg 1\),
ensuring consistency with static scaling \(A_{\infty} \sim (g_c - g)^\beta\)~\cite{henkel2000}.

\begin{figure*}[]
\centering
  \includegraphics[width=0.45\linewidth]{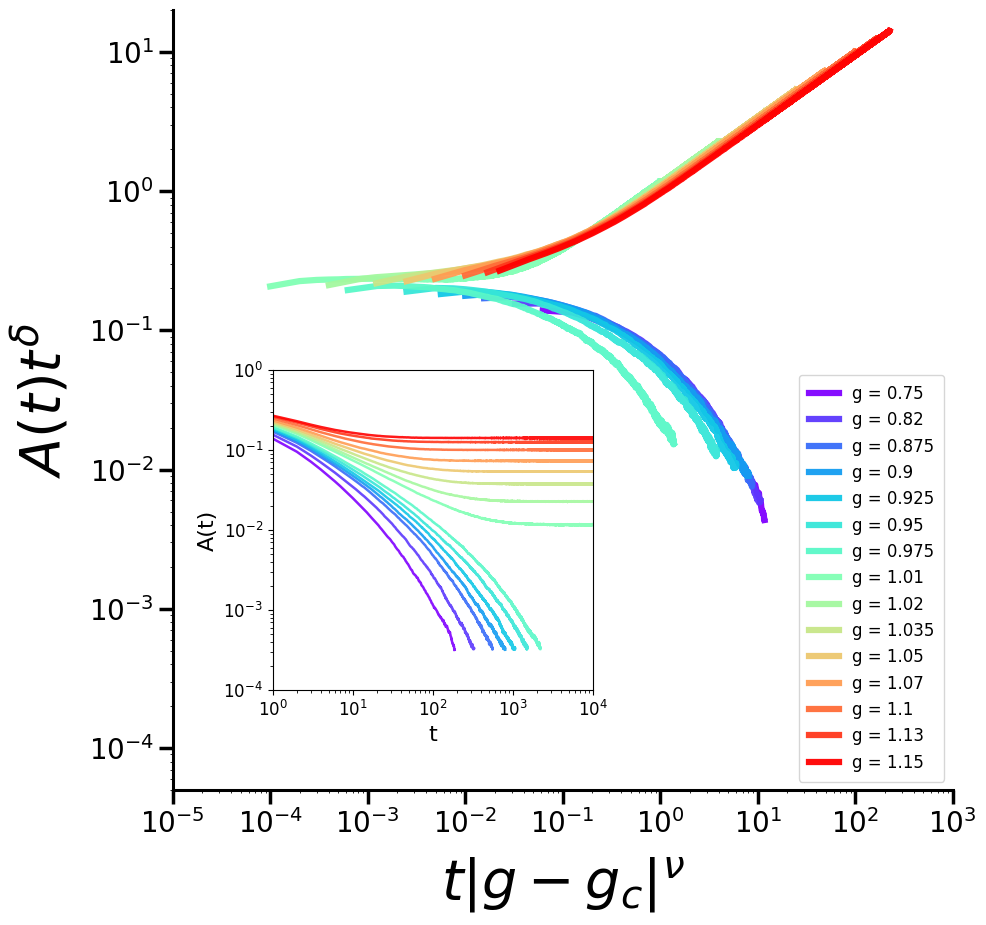}
  \includegraphics[width=0.45\linewidth]{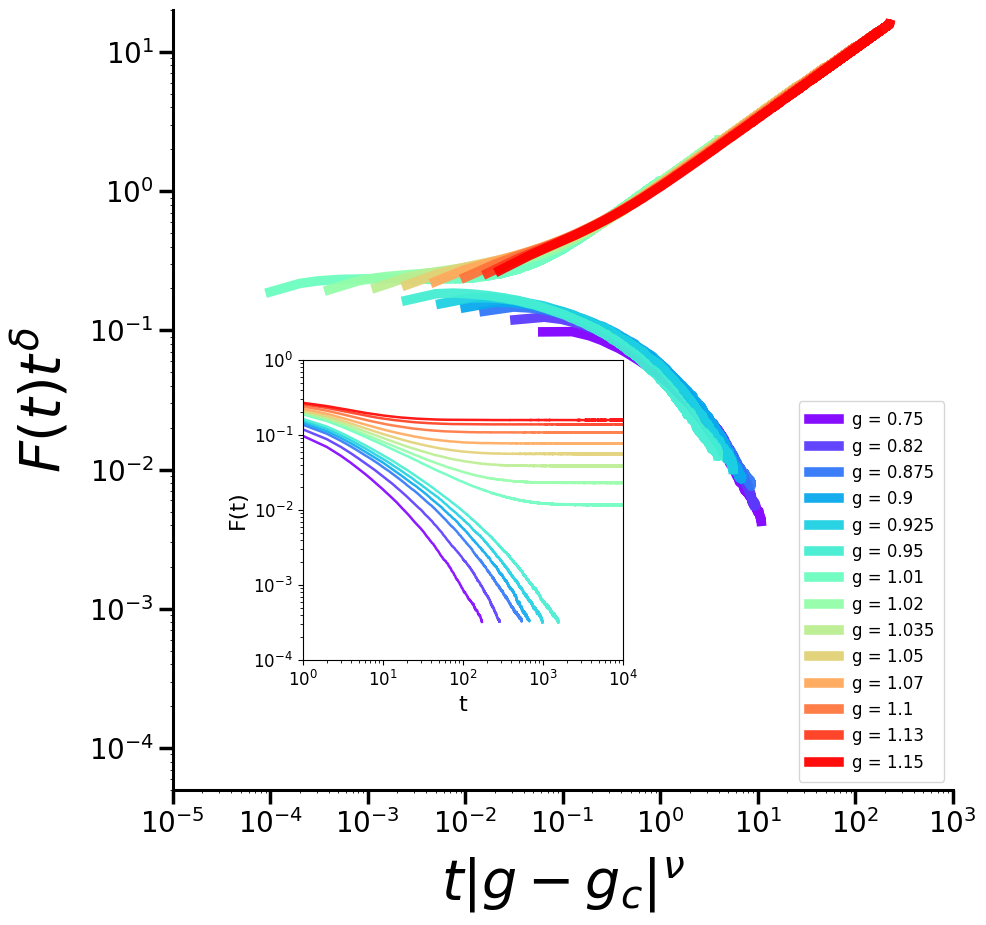}
\caption{Dynamical scaling collapses for instantaneous strategy (strategy A) using mean field exponents \((\delta,\nu_{\parallel})=(0.5,2.0)\). Data  Collapse shows superposition into master curve across \(g\) values. This confirms single-parameter finite-time scaling and mean field universality~\cite{vemula2025}.}
\label{fig:instantaneous_collapse}
\end{figure*}

We test this dynamical scaling for strategy A, which should conform to the mean field
predictions. The data collapse procedure involves plotting \(A(t)t^\delta\) against \(t|g-g_c|^{\nu_{\parallel}}\)
and assessing the quality of superposition across different \(g\) values, using rainbow-colormapped curves for \(g \in [0.75,1.15]\) (excluding \(g_c\)) (see Fig.~\ref{fig:instantaneous_collapse}).

% INSTANTANEOUS COLLAPSE FIGURE

The data collapses for strategy A indicate consistency with the
predicted exponents \(\delta=0.5\) and \(\nu_{\parallel}=2.0\). This confirms that the critical behavior corresponding to strategy A belong to the
mean field universality class, as expected from theoretical
considerations~\cite{henkel2008}. 

% =====================================================================
%\section{Threshold Dynamics and Effects}
%\label{sec:threshold}
% =====================================================================
Threshold-based decision rules (strategy B) introduce fundamental differences from instantaneous rules. Unlike instantaneous rules where
decisions depend only on current population differences, threshold rules require agents
to look at the average population (ignoring the majority/minority status) before switching states. 

% COMBINED THRESHOLD COLLAPSE FIGURE - UP AND DOWN VARIANTS
\begin{figure*}[]
\centering
  \includegraphics[width=0.45\linewidth]{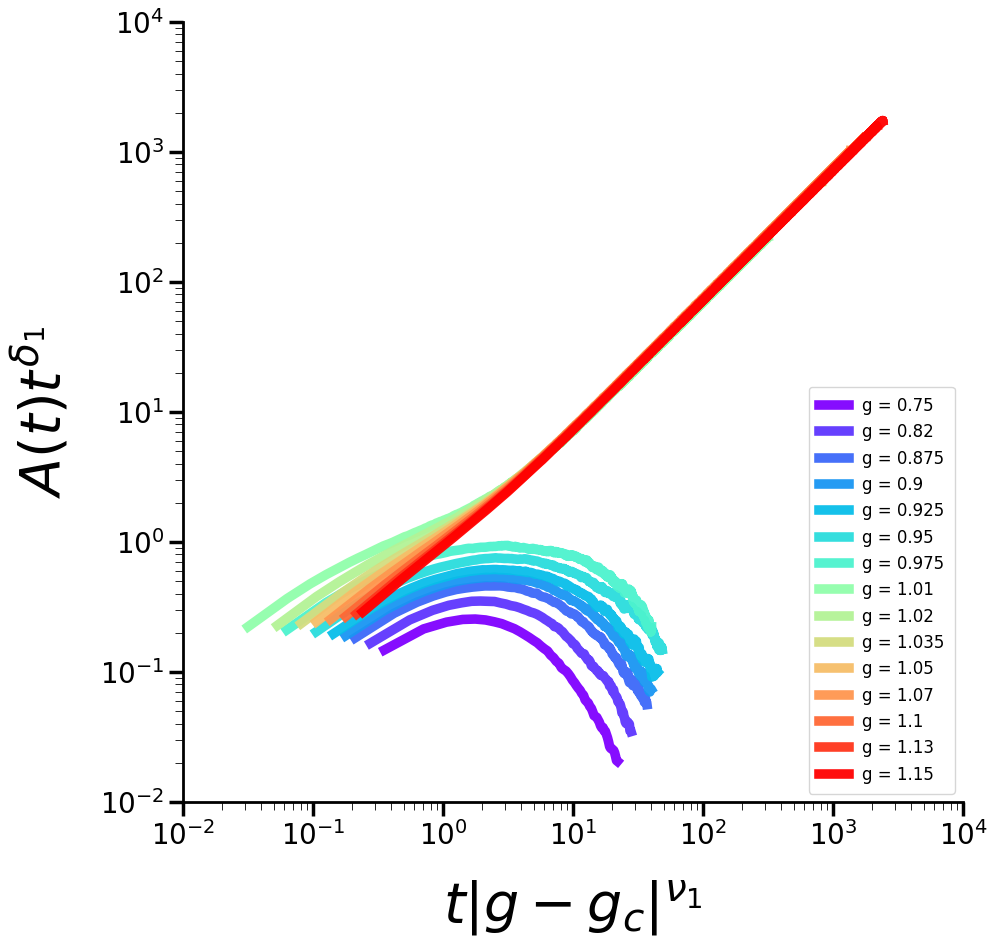}
  \includegraphics[width=0.45\linewidth]{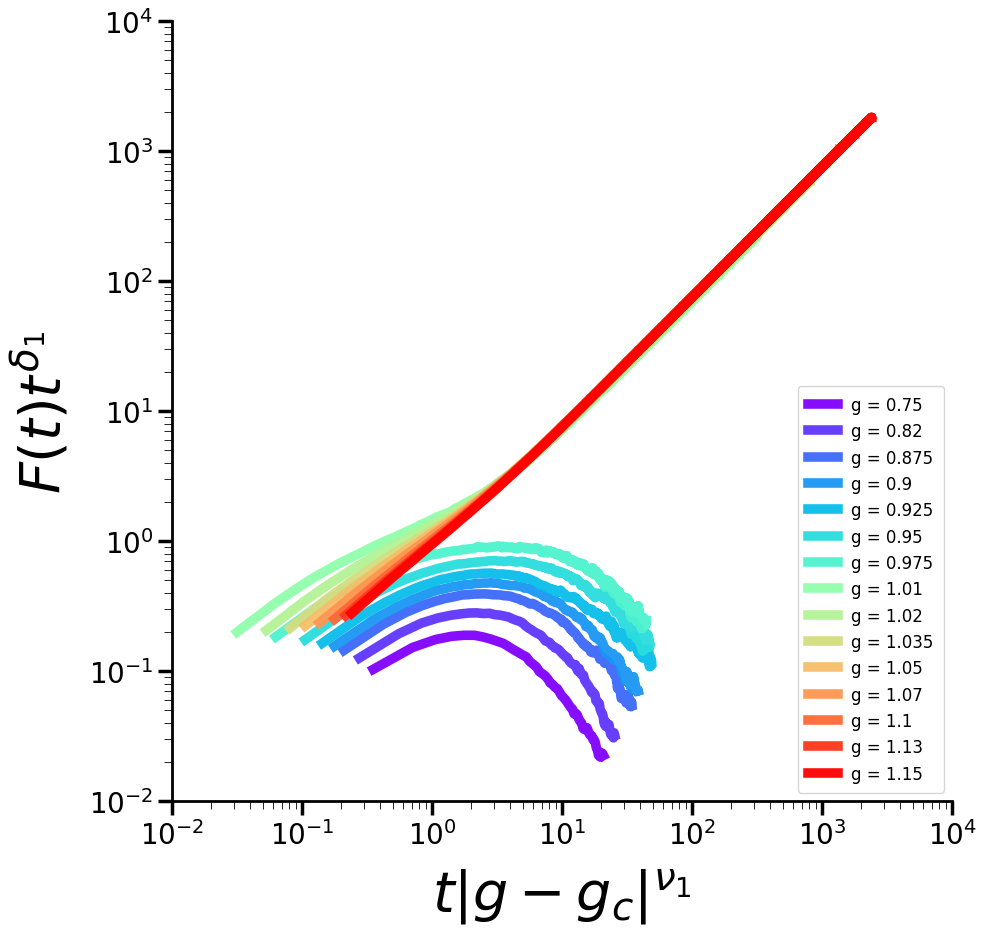}
  \includegraphics[width=0.45\linewidth]{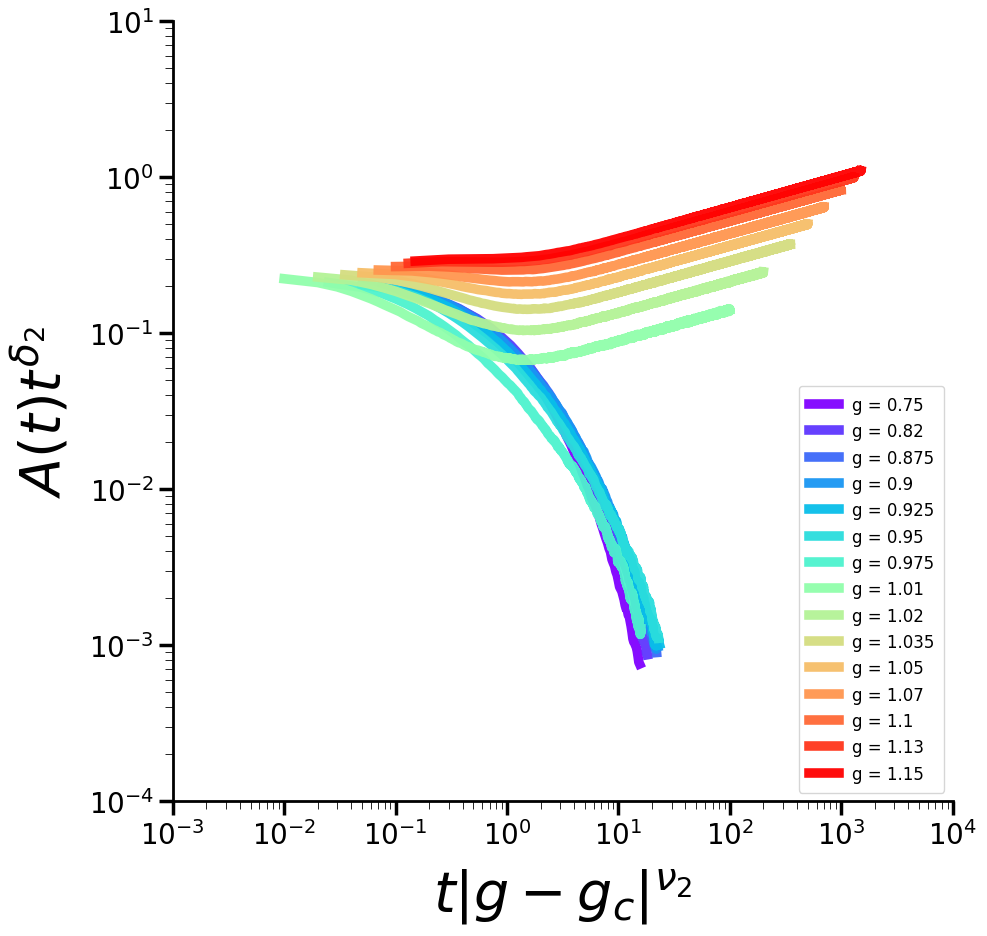}
  \includegraphics[width=0.45\linewidth]{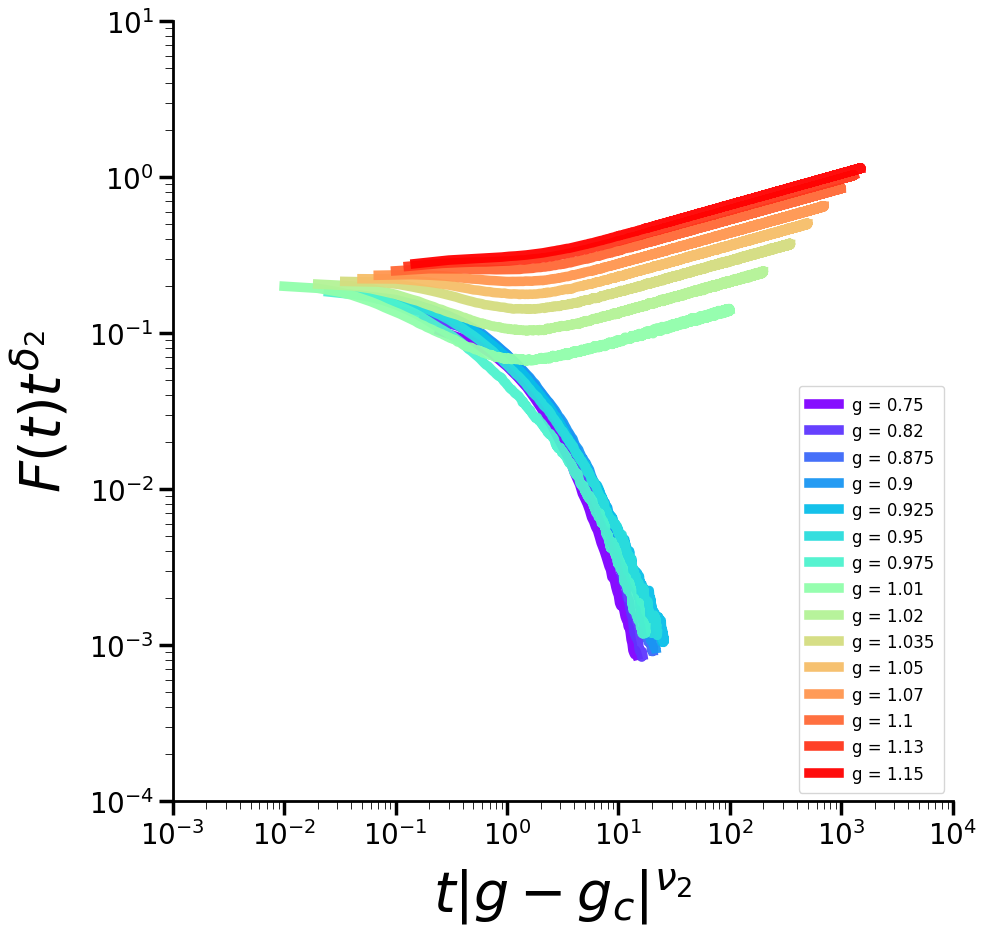}
\caption{Dynamical scaling collapses for threshold strategies (strategy B) using non-mean field exponents. (a-b) Supercritical regime: \((\delta,\nu_{\parallel})=(1.0,0.75)\). (c-d) Subcritical regime: \((\delta,\nu_{\parallel})=(0.2,1.0)\). Log-log plots reveal systematic fanning when using mean field exponents, with partial up/down overlaps indicating multiple timescales. Failure across variants and observables confirms deviation from mean field universality, suggesting a distinct universality class.}
\label{fig:threshold_collapse}
\end{figure*}

The deviations (from strategy A and mean field expectations) are apparent in the attempts to data collapse (see Fig.~\ref{fig:threshold_collapse}). Attempts to
collapse the data using mean field exponents produce systematic curvature and poor superposition.
We find different scaling regimes: the supercritical region requires exponents \(\delta=1.0\) and \(\nu_{\parallel}=0.75\), while the subcritical region requires \(\delta=0.2\) and \(\nu_{\parallel}=1.0\). This persistent failure, combined with the consistently different static exponent
\(\beta \approx 0.75\), provides compelling evidence for a distinct universality class for the model following strategy B.

% =====================================================================
\section{Discussions and conclusions}
\label{sec:discussion}
% =====================================================================

%The threshold dynamics in PMG provide a bridge between economics-inspired agent models and physical systems, suggesting universal mechanisms by which decision rules alter collective criticality. This connects with earlier work on adaptive minority games that demonstrated how genetic algorithm based adaptation could drive systems toward different behavioral regimes through threshold mechanisms~\cite{sysiaho2004adaptive}.

%The common practice of classifying active-absorbing transitions based solely on symmetry and conservation laws may be insufficient for systems with threshold-based decision rules. Our work demonstrates that microscopic rules with threshold behavior can create new universality classes, necessitating more nuanced classification schemes for complex adaptive systems.

%Researchers should exercise caution when mapping agent-based models to established universality classes. Even qualitative similarity in macroscopic behavior (active-absorbing transitions) does not guarantee universality if microscopic rules introduce threshold behavior or other relevant perturbations.

% =====================================================================
%\section{Conclusions and Future Directions}
%\label{sec:conclusions}
% =====================================================================
We have presented a comprehensive numerical study of phase transitions in the Parallel Minority Game under two distinct decision paradigms. Our key findings are that instantaneous decision rules (strategy A) yield an active-absorbing transition in terms of the fraction of agents not in the minority or the fraction of sites where over-crowding happens, with critical exponents (\(\beta \approx 0.95\), \(\delta \approx 0.5\), \(\nu_{\parallel} \approx 2.0\)) (see Fig.~\ref{fig:beta_group} and ~\ref{fig:instantaneous_collapse}) and consistent with mean field values for such exponents. But the threshold-based rules (strategy B) produce a distinct universality class characterized by \(\beta \approx 0.75\) (see Fig.~\ref{fig:beta_group}) and systematic failure of dynamical scaling, with evidence of multi-timescale relaxation and different exponent values in supercritical (\(\delta=1.0\), \(\nu_{\parallel}=0.75\)) (see Fig.~\ref{fig:threshold_collapse}) (a-b)  and subcritical (\(\delta=0.2\), \(\nu_{\parallel}=1.0\)) (see Fig.~\ref{fig:threshold_collapse}) (c-d) regimes. The deviation stems from the fundamental nature of thresholding, which acts as a relevant perturbation in the renormalization group sense.

We should also note the possible link of the model dynamics to a frustrated spin-glass like system \cite{sk_model}. In such systems, due to both ferromagnetic and anti-ferromagnetic coupling between (Ising) spins, the ground state becomes highly degenerate and difficult to find, the energy landscape being highly corrugated. The system dynamics can often lead to a local minimum rather than the global one. Similarly, in the present system, decrease in population in one site necessarily increases the same in other sites, making the 'ground state' or optimized distribution of crowds (presumably equal population in each site) a difficult goal. The effect of such 'frustration' in the critical exponent values studied above, and the effectiveness of heuristic techniques (see e.g., \cite{sg_pre}) to find the ground states here could be an important direction. 

In conclusion, our results demonstrate that minimal cognitive features such as decision thresholds can fundamentally alter universality classes in adaptive multi-agent systems. This has several important implications.

Real-world decision makers often exhibit inertia and threshold behavior-consumers wait for significant price changes before switching brands, investors require substantial evidence before changing portfolios, and firms delay responses until profit differentials exceed certain thresholds. Our findings suggest that such micro-level cognitive features can dramatically affect macroscopic collective behavior and phase transition characteristics.

\section*{Code Availability}
The simulation code for this work is available at
\href{https://github.com/AryanTyagi-Complexity/Parallel-Minority-Game}{https://github.com/AryanTyagi-Complexity/Parallel-Minority-Game}

\bibliography{aipsamp}% Produces the bibliography via BibTeX.

@article{chakraborti2011econophysicsII,
  title        = {Econophysics review: II. Agent‐based models},
  author       = {Chakraborti, Anirban and Muni Toke, Ioane and Patriarca, Marco and Abergel, Fr{\'e}d{\'e}ric},
  journal      = {Quantitative Finance},
  volume       = {11},
  number       = {7},
  pages        = {1013--1041},
  year         = {2011},
  doi          = {10.1080/14697688.2010.539249},
  publisher    = {Taylor \& Francis}
}

@book{sociophysics2014,
  author    = {Sen, Parongama and Chakrabarti, Bikas K.},
  title     = {Sociophysics: An Introduction},
  publisher = {Oxford University Press},
  address   = {Oxford, UK},
  year      = {2014},
  isbn      = {9780199662456}
}

@article{challet1997,
  author  = {Challet, D. and Zhang, Y.-C.},
  title   = {Emergence of cooperation and organization in an evolutionary game},
  journal = {Physica A},
  volume  = {246},
  pages   = {407--418},
  year    = {1997},
  doi     = {10.1016/S0378-4371(97)00295-6}
}

@article{challet1998,
  author  = {Challet, D. and Zhang, Y.-C.},
  title   = {On the minority game: Analytical and numerical studies},
  journal = {Physica A},
  volume  = {256},
  pages   = {514--532},
  year    = {1998},
  doi     = {10.1016/S0378-4371(98)00470-7}
}

@article{challet1999,
  author  = {Challet, D. and Marsili, M.},
  title   = {Phase transition and symmetry breaking in the minority game},
  journal = {Phys. Rev. E},
  volume  = {60},
  pages   = {R6271--R6274},
  year    = {1999},
  doi     = {10.1103/PhysRevE.60.R6271}
}

@article{arthur1994,
  author  = {Arthur, W. Brian},
  title   = {Inductive reasoning and bounded rationality},
  journal = {Am. Econ. Rev.},
  volume  = {84},
  pages   = {406--411},
  year    = {1994}
}

@book{challet2005,
  author    = {Challet, D. and Marsili, M. and Zhang, Y.-C.},
  title     = {Minority Games},
  publisher = {Oxford Univ. Press},
  address   = {Oxford},
  year      = {2005}
}

@article{chakraborti2015,
  author  = {Chakraborti, A. and Challet, D. and Chatterjee, A. and Marsili, M. and Zhang, Y.-C. and Chakrabarti, B. K.},
  title   = {Statistical mechanics of competitive resource allocation using agent-based models},
  journal = {Phys. Rep.},
  volume  = {552},
  pages   = {1--25},
  year    = {2015},
  doi     = {10.1016/j.physrep.2014.09.004}
}

@article{ho2004,
  author  = {Ho, K. H. and Chow, F. K. and Chau, H. F.},
  title   = {Wealth inequality in the minority game},
  journal = {Phys. Rev. E},
  volume  = {70},
  pages   = {066110},
  year    = {2004},
  doi     = {10.1103/PhysRevE.70.066110}
}

@article{zhang2016,
  author  = {Zhang, J. and Huang, Z. and Wu, Z. and Su, R. and Lai, Y.-C.},
  title   = {Controlling herding in minority game systems},
  journal = {Sci. Rep.},
  volume  = {6},
  pages   = {20925},
  year    = {2016},
  doi     = {10.1038/srep20925}
}

@article{challet2001local,
  author  = {Challet, D. and Chessa, A. and Marsili, M. and Zhang, Y.-C.},
  title   = {From minority games to real markets},
  journal = {Quant. Finance},
  volume  = {1},
  pages   = {168--176},
  year    = {2001},
  doi     = {10.1088/1469-7688/1/2/303}
}

@article{huang2012,
  author  = {Huang, Z. and Zhang, J. and Dong, J. and Huang, L. and Lai, Y.},
  title   = {Emergence of grouping in multi-resource minority game dynamics},
  journal = {Sci. Rep.},
  volume  = {2},
  pages   = {703},
  year    = {2012},
  doi     = {10.1038/srep00703}
}

@article{bottazzi2007,
  author  = {Bottazzi, G. and Devetag, G.},
  title   = {Competition and coordination in experimental minority games},
  journal = {J. Evol. Econ.},
  volume  = {17},
  pages   = {241--275},
  year    = {2007},
  doi     = {10.1007/s00191-006-0041-6}
}

@article{dhar2011,
  author  = {Dhar, D. and Sasidevan, V. and Chakrabarti, B. K.},
  title   = {Emergent cooperation amongst competing agents in minority games},
  journal = {Physica A},
  volume  = {390},
  pages   = {3477--3485},
  year    = {2011},
  doi     = {10.1016/j.physa.2011.05.040}
}

@article{chakrabarti2009,
  author  = {Chakrabarti, A. S. and Chakrabarti, B. K. and Chatterjee, A. and Mitra, M.},
  title   = {The Kolkata Paise Restaurant problem and resource utilization},
  journal = {Physica A},
  volume  = {388},
  pages   = {2420--2426},
  year    = {2009},
  doi     = {10.1016/j.physa.2009.02.037}
}

@article{ghosh2010,
  author  = {Ghosh, A. and Chatterjee, A. and Chakrabarti, B. K. and Mitra, M.},
  title   = {Statistics of the Kolkata Paise Restaurant problem},
  journal = {New J. Phys.},
  volume  = {12},
  pages   = {075073},
  year    = {2010},
  doi     = {10.1088/1367-2630/12/7/075073}
}

@article{biswas2012,
  author  = {Biswas, S. and Ghosh, A. and Chatterjee, A. and Naskar, T. and Chakrabarti, B. K.},
  title   = {Continuous transition of social efficiencies in the stochastic-strategy minority game},
  journal = {Phys. Rev. E},
  volume  = {85},
  pages   = {031104},
  year    = {2012},
  doi     = {10.1103/PhysRevE.85.031104}
}

@article{biswas2021,
  author  = {Biswas, S. and Mandal, A. K.},
  title   = {Parallel minority game and its application in movement optimization during an epidemic},
  journal = {Physica A},
  volume  = {561},
  pages   = {125271},
  year    = {2021},
  doi     = {10.1016/j.physa.2020.125271}
}

@article{henkel2000,
  author  = {Henkel, M. and Hinrichsen, H. and L{\"u}beck, S.},
  title   = {Non-equilibrium critical phenomena: Phase transitions into absorbing states},
  journal = {Adv. Phys.},
  volume  = {49},
  pages   = {815--958},
  year    = {2000},
  doi     = {10.1080/00018730050198152}
}

@article{odor2004,
  author  = {{\'O}dor, G.},
  title   = {Universality classes in nonequilibrium lattice systems},
  journal = {Rev. Mod. Phys.},
  volume  = {76},
  pages   = {663--724},
  year    = {2004},
  doi     = {10.1103/RevModPhys.76.663}
}

@article{lubeck2004,
  author  = {L{\"u}beck, S.},
  title   = {Universal scaling behavior of non-equilibrium phase transitions},
  journal = {Int. J. Mod. Phys. B},
  volume  = {18},
  pages   = {3977--4118},
  year    = {2004},
  doi     = {10.1142/S0217979204025998}
}

@article{sysiaho2004adaptive,
  author  = {Sysi-Aho, M. and Chakraborti, A. and Kaski, K.},
  title   = {Searching for good strategies in adaptive minority games},
  journal = {Phys. Rev. E},
  volume  = {69},
  pages   = {036125},
  year    = {2004},
  doi     = {10.1103/PhysRevE.69.036125}
}

@article{moelbert2002local,
  author  = {Moelbert, S. and De Los Rios, P.},
  title   = {The local minority game},
  journal = {Physica A},
  volume  = {303},
  pages   = {217--225},
  year    = {2002},
  doi     = {10.1016/S0378-4371(01)00627-4}
}

@misc{vemula2025,
  author       = {Vemula, A. R. and Biswas, S.},
  title        = {Efficient Strategy for Parallel Minority Games},
  howpublished = {arXiv:2509.02770},
  year         = {2025},
  eprint       = {2509.02770},
  archivePrefix= {arXiv},
  primaryClass = {physics.soc-ph}
}

@book{henkel2008,
  author    = {Henkel, Malte and Hinrichsen, Haye and L{\"u}beck, Sven},
  title     = {Non-Equilibrium Phase Transitions: Volume 1 -- Absorbing Phase Transitions},
  publisher = {Springer},
  year      = {2008},
  address   = {Dordrecht}
}

@book{chakrabarti2006econophysics,
  title={Econophysics and sociophysics: trends and perspectives},
  author={Chakrabarti, Bikas K and Chakraborti, Anirban and Chatterjee, Arnab},
  year={2006},
  publisher={John Wiley \& Sons}
}

@book{chakrabarti2023data,
  title={Data science for complex systems},
  author={Chakrabarti, Anindya S and Bakar, K Shuvo and Chakraborti, Anirban},
  year={2023},
  publisher={Cambridge University Press}
}

@article{pmg_vac,
	author={Xue, Chenli and Luo, Xiaofeng and Sun, Gui-Quan},
	title={Vaccination-Transmission Coupled Mechanism Based on Parallel Minority Game},
	journal={Chinese Physics B},
	url={http://iopscience.iop.org/article/10.1088/1674-1056/ae306a},
	year={2025}
}

@article{sg_pre,
  title = {Classical annealing of the {S}herrington-{K}irkpatrick spin glass using {S}uzuki-{K}ubo mean-field Ising dynamics},
  author = {Das, Soumyaditya and Biswas, Soumyajyoti and Chakrabarti, Bikas K.},
  journal = {Phys. Rev. E},
  volume = {112},
  issue = {1},
  pages = {014104},
  numpages = {7},
  year = {2025},
  month = {Jul},
  publisher = {American Physical Society},
  doi = {10.1103/www8-3ts1},
  url = {https://link.aps.org/doi/10.1103/www8-3ts1}
}

@article{sk_model,
  title = {Solvable Model of a Spin-Glass},
  author = {Sherrington, David and Kirkpatrick, Scott},
  journal = {Phys. Rev. Lett.},
  volume = {35},
  issue = {26},
  pages = {1792--1796},
  numpages = {0},
  year = {1975},
  month = {Dec},
  publisher = {American Physical Society},
  doi = {10.1103/PhysRevLett.35.1792},
  url = {https://link.aps.org/doi/10.1103/PhysRevLett.35.1792}
}

\end{document}